\documentstyle[prb,aps]{revtex}
\input epsfig.sty

\begin{document}

\advance\textheight by 0.2in
\twocolumn[\hsize\textwidth\columnwidth\hsize\csname@twocolumnfalse%
\endcsname

\draft
\title{ { \it to appear in Phys. Rev. B } \\
Edge effects in a frustrated Josephson junction array with modulated
couplings}
\author{E. Granato$^{1}$, J.M. Kosterlitz$^{2}$ and M.V. Simkin$^{2}$}
\address{$^1$ Laborat\'{o}rio Associado de Sensores e Materiais, Instituto Nacional
de Pesquisas Espacias, 12201 S\~{a}o Jos\'{e} dos Campos, SP, Brazil. \\
$^{2}$ Department of Physics, Brown University, Providence, RI 02912, USA}
\maketitle

\begin{abstract}
A square array of Josephson junctions with modulated strength in a magnetic
field with half a flux quantum per plaquette is studied by analytic
arguments and dynamical simulations. The modulation is such that alternate
columns of junctions are of different strength to the rest. Previous work
has shown that this system undergoes an $XY$ followed by an Ising-like
vortex lattice disordering transition at a {\it lower} temperature. We argue
that resistance measurements are a possible probe of the vortex lattice
disordering transition as the linear resistance $R_{L}(T)\sim A(T)/L$ with $
A(T) \propto (T-T_{cI})$ at intermediate temperatures $T_{cXY}>T>T_{cI}$ due
to dissipation at the array edges for a particular geometry and vanishes for
other geometries. Extensive dynamical simulations are performed which
support the qualitative physical arguments.
\end{abstract}


]

There has been a lot of interest in arrays of superconducting grains coupled
by Josephson junctions. In the absence of a magnetic field perpendicular to
the plane of the array, such a system is rather well described by a
classical two dimensional $XY$ model and experimental transport measurements
\cite{martinoli} are well described by the dynamical extension\cite{ahns} of
the static theory\cite{kt,jmk}. The most recent technique employed to
measure truly equilibrium properties of a superconducting array is a
magnetic flux noise measurement\cite{shaw} which does not require an imposed
current. Earlier experimental studies measured the voltage $V$ due to an
applied zero frequency current $I$ \cite{dc1,dc2,dc3,dc4,dc5,dc6,zant} and
the frequency dependent impedance $Z(\omega )$ by two coil mutual inductance
techniques \cite{martinoli}. The agreement between theory and experiment is
quite good. In the presence of a magnetic field normal to the array, the
situation is not so clear as, even in the simplest case of half a flux
quantum per elementary plaquette $f=1/2$ on a square lattice, the system is
much more complicated and less well understood. Even in the absence of
disorder, which is inevitably present in an experimental system\cite{zant},
there are two types of competing order when $f=1/2$: a discrete $Z_2$
symmetry of the ground state of the vortex lattice and the $U(1)$ symmetry
of the superconducting order parameter. In an isotropic system where all the
junction strengths are the same the phase coherence of the superconducting
order parameter must be destroyed at a lower temperature or, at best, the
same temperature as the discrete order of the vortex lattice \cite{gkn}. The
most recent simulations on this\cite{ols,lee} agree that the phase order is
destroyed by a $KT$ transition at a very slightly lower temperature than the
discrete order which undergoes a transition in the Ising universality class.

An interesting variant of the isotropic square array with $f=1/2$ was
proposed by Berge {\it et al}\cite{berge} in which alternate columns of
junctions are of different strength to the rest. This model is described by
the Hamiltonian 
\begin{equation}
-H/k_BT=\sum_{<{\bf r,r^{\prime }}>}K_{{\bf r,r^{\prime }}}\cos [\theta
({\bf r})-\theta ({\bf r^{\prime }})-A({\bf r,r^{\prime }})]  \label{eq:1}
\end{equation}
where $K_{{\bf r,r^{\prime }}}=K=J/k_BT$ on all nearest neighbor bonds
except on alternating columns where $K_{{\bf r,r+\hat{y}}}=\eta K$. $A({\bf
r,r^{\prime }})=\frac{2\pi }{\Phi _0}\int_{{\bf r}}^{{\bf r^{\prime }}}{\bf 
A\cdot dl}$ where ${\bf A}$ is the vector potential of the external magnetic
field and $\Phi _0$ is the quantum of flux. The frustration of plaquette $
{\bf R}$ is $f({\bf R})=\sum_{{\bf \Box R}}A({\bf r,r^{\prime }})/2\pi =1/2$
where the sum is over the bonds $<{\bf r,r^{\prime }}>$ surrounding the
plaquette. It is a gauge invariant definition and a convenient gauge is the
Landau gauge in which $A({\bf r,r^{\prime }})=\pi $ on every second column
of bonds and $A({\bf r,r^{\prime }})=0$ otherwise. Such an array is
experimentally realizable  by varying the area of the appropriate
junctions to obtain alternating columns of strength $\eta K$ and is 
currently being investigated \cite{pm}. The model
described by Eq.(\ref{eq:1}) has been studied numerically\cite
{berge,eikmans,gk1} and it was found that, by varying the anisotropy
parameter $\eta $, the order of the $XY$ and Ising transition is reversed.
In the isotropic system at $\eta =1$ the $XY$ transition occurs at a lower
temperature than the Ising transition but when $|\eta -1|$ is sufficiently
large the Ising is at a lower temperature than the $XY$ transition. The
early simulations were unable to separate the transitions for $|\eta
-1|\approx 0$ but later work\cite{ols,lee} provided convincing evidence for
two separate transitions with $T_{cI}>T_{cXY}$. Another generalized version
of the square array at $f=1/2$ has been considered recently \cite{bg} in
which the couplings $K_{r,r+\hat{y}}$ in Eq. (\ref{eq:1}) are modulated in
the $x$ and $y$ directions simultaneously, in a zig-zag pattern. In this
case, varying $\eta $ in the range $\eta >0.6$ does not lead to a change in
the order of the $Z_2$ and $U(1)$ transitions and they critical properties
remain the same as for the isotropic square array.

In a two dimensional plane of superconductor in a magnetic field normal
to the plane a vortex lattice is formed. If a vortex is subjected to a
force due to an external applied current, it will move in a direction
perpendicular to the current and this creates a voltage $V \sim n_{f}I$
where $n_{f}$ is the density of free vortices, which implies a finite
linear resistance. This is regarded as signalling the destruction of
superconductivity. An array of Josephson junctions with no disorder in
a magnetic field with a frustration $f=p/q$ has a ground state which
is a vortex lattice commensurate with the underlying lattice and is
pinned and superconducting. As the temperature is raised, the vortex
lattice will melt or become a floating phase which is not pinned. In
either case, one would naively expect that the system becomes non
superconducting as the vortices will move and induce a voltage when an
external current is applied. In particular, a square array with $f=1/2$
and $\eta >1/3$ has a ground state \cite{berge} which is a vortex lattice
with a vortex in every second plaquette commensurate with the underlying
lattice. Since the vortex lattice is pinned by commensurability effects,
the system will be superconducting and should become non-superconducting
when the lattice melts. In the anisotropic array of Eq.(\ref{eq:1}), the
vortex lattice melts by an Ising transition at $T=T_{cI}$ but $XY$ order
(superconductivity) persists to higher temperatures \cite{berge,eikmans}
although the vortex lattice has melted, which seems to contradict the
standard folklore that, when the vortex lattice melts, superconductivity
disappears. It seems to us that there are three possibilities to reconcile
the equilibrium behavior of the anisotropic system of Eq.(\ref{eq:1})
with these qualitative arguments about the transport properties with a
small applied current $I$ at intermediate temperature $T_{cI}<T<T_{cXY}$
where the vortex lattice is melted: (1) the present understanding of the
effect of the melting of a vortex lattice on superconductivity is wrong;
(2) the equilibrium calculations at $I=0$ have no relation to the dynamics
as $I\rightarrow 0$; (3) the properties of the melted vortex lattice
in the anisotropic array are special and the qualitative argument that
superconductivity is destroyed by the melting does not apply.

In this paper, we take the view that the system of Eq.(\ref{eq:1}) is
special and that scenario (3) is the explanation and also address the
question of the signature (if any) of the low temperature
Ising transition in experimental measurements\cite{pm} on an anisotropic
array with column modulation. Two coil mutual inductance experiments measure
the dynamical impedance $Z(\omega )$ which is proportional to the inverse of
the helicity modulus \cite{martinoli} $\gamma $. At the Ising transition,
this has a harmless singularity of the form\cite{eikmans,gk1} $\gamma \sim
t\ln t$ where $t=T/T_{cI}-1$ which implies that the impedance will also not
show any divergence as observable signature. The implications for the flux
noise spectrum have not been worked out and it would be of some interest to
do this. There is one possibility which is the linear resistance which we
discuss in the rest of this paper. We study by qualitative analytical
methods and by numerical simulations the zero frequency $IV$ characteristics
of an anisotropic array and show that the onset of linear resistance occurs
at the low temperature Ising transition, which may be detectable by
experiment. This is an edge effect and, in an $L\times L$ array, the linear
resistance $R_L(T)\sim A(T)/L$ with $A(T)>0$ when $T>T_{cI}$. We begin by
defining the model and argue that the onset of linear resistance is due to
the formation and growth of domains of reversed chiral order at the edges of
the array. The geometry of the array and the direction of the applied
current is important, and we first define the various geometries and
boundary conditions and what is to be expected on physical grounds in each
case. We then discuss the results of numerical simulations and show that
these are consistent with our analytical arguments.

The low temperature Ising transition may be described in terms of the
proliferation of domain walls separating regions of different ground states.
A ground state configuration of Eq.(\ref{eq:1}) may be represented in the
Coulomb gas representation by a set of unit charges in a checkerboard
pattern on half the sites of the dual lattice\cite{eikmans}. The important
low energy excitations which destroy the Ising order are domain walls
between regions of different Ising order and these domain walls lie along
the bonds of the original lattice. It is well known that there are
fractional charges $q=\pm 1/4$ at the corners of these domain walls\cite
{halsey,korsh,thij}. One may argue that such domains with a net integer
charge formed of corner charges of the same sign are the excitations which
undergo a $KT$ unbinding transition at the same temperature as the Ising
transition. However, numerical simulations\cite{ols,lee,berge,eikmans} show
that the $XY$ and Ising transitions are separated for almost all values of
the anisotropy parameter $\eta $ which implies that the fractional corner
charges and the integer charges do not screen each other. In the case of
interest here, $|\eta -1|$ sufficiently large, one can understand this by
considering the domain walls at $T_{cI}<T<T_{cXY}$. As shown by Eikmans {\it 
et al.}\cite{eikmans}, the domain walls between regions of different Ising
ground states lie along the strong bonds since the energy of these is less
than those on the weak bonds. In fact, they find numerically that there is a
negligible density of domain walls on weak bonds for $T<T_{cXY}$ but the
density on the strong bonds grows rapidly when $T>T_{cI}$. Thus, these
domains carry zero net charge as the fractional corner charges must
alternate between $q=\pm 1/4$. For a domain to have an integer charge, one
of the vertical walls must lie along a weak bond which costs a lot of free
energy. The unit cell describing the Ising degree of freedom may be taken to
be four adjacent elementary plaquettes bounded by strong bonds as shown in
Fig. (1) and at $T\gtrsim T_{cI}$ domains with an integer number of these
unit cells will proliferate. Since these domains carry zero net charge and
zero dipole moment, they cannot contribute to the linear resistance which is
proportional to $n_f$ , the concentration of thermally excited free charges
(vortices)\cite{ahns}. Because of the zero dipole moment, the mechanism of
stretching of domains by the applied current $I$ suggested by Mon and Teitel
\cite{mt} as a contribution to the non linear resistance for $T<T_{cI}$ of
the form $V\sim I\exp (-f_d(T)/IkT)$, where $f_d(T)$ is the domain wall free
energy, is not a dominant effect for the anisotropic arrays considered in this paper.

\begin{figure}[tbp]
\centering\epsfig{file=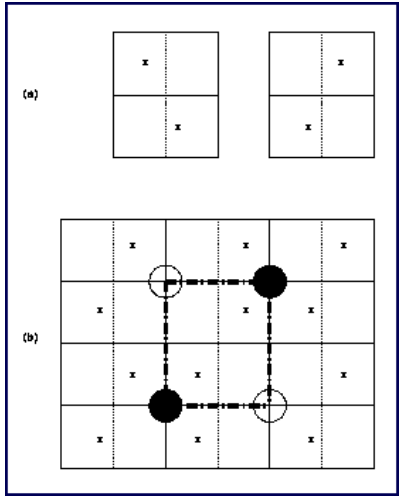,bbllx=0.5cm,bblly=7cm,bburx=18cm,
bbury=22cm,angle=-90,width=8cm}
\caption{(a) States of opposite Ising order of $2\times 2$ unit cells.
Solid(dotted) lines denote strong(weak) bonds. Unit charges are represented
by x's. (b) A domain of opposite Ising order. The domain wall is the
dash-dotted line. Filled(unfilled) dots at domain wall corners are $\pm 1/4$
fractional charges.}
\label{fig1}
\end{figure}

We consider a $2N_x\times 2N_y$ lattice of elementary plaquettes
corresponding to a $(2N_x+1)\times (2N_y+1)$ lattice of superconducting
grains coupled by weak links. There are two geometries to consider (a)
columns of weak bonds at the two vertical edges and (b) columns of strong
bonds at the edges as sketched in Figs.(2a) and (2b). In case (b), there are 
$N_x\times N_y$ complete unit cells and the current is injected uniformly in
the $x$ direction perpendicular to the weak bonds. A linear resistance is
due to the current driving the pre-existing thermally excited charges across
the system but such charges do not exist in the bulk because, near the Ising
transition, only neutral domains are present as the domain walls are
constrained to lie along the strong bonds\cite{eikmans}. One may expect that
free, unbound fractional charges exist near the edges of the array because
of domain formation at the edges. We expect that for $T\gtrsim T_{cI}$ the
system will consist of a set of domains of opposite Ising order of size
determined by the bulk correlation length $\xi _b\sim A_bt^{-\nu _b}$ in the
bulk and at the edges for free boundary conditions by the surface
correlation length $\xi _{\Vert }\sim A_{\Vert }t^{-\nu _{\Vert }}$ with $
\nu _b=\nu _{\Vert }=1$\cite{mcoy}. The amplitudes $A_b$ and $A_{\Vert }$
are nonuniversal but their ratio is universal and is given in terms of
critical exponents from conformal invariance\cite{cardy} by 
\begin{equation}
A_{\Vert }/A_b=\eta _b/\eta _{\Vert }=1/4  \label{eq:2}
\end{equation}
with the bulk and edge exponents \cite{mcoy,binder,cardy} $\eta _b=1/4$ and $
\eta _{\Vert }=$ $1$ which implies that the density of edge domains is
larger than that in the bulk of the array. This is of no significance for
the linear resistance in arrays where the strong bonds are at the edges as
in Fig. (2b) as there are no net fractional charges at the ends of these
edge domains and these will behave just like the domains in the bulk of Fig.
(1b) which carry no net charge. The situation for arrays with weak bonds at
the vertical edges is very different and this is shown in Fig. (2a). In this
geometry, one can regard these edges as splitting the unit cells of Fig.
(2a) in half thus forcing fractional charges of opposite signs at the two
ends of the domains. Since $T>T_{cI}$, the domain walls may be regarded as
having melted and having zero line tension as the edge undergoes an ordinary
transition slaved to the bulk transition\cite{binder}. Along the edges of
the array of linear size $L$, these charges are effectively free unbound
charges of concentration $n_f\sim 1/L\xi _{\Vert }$. This implies that in an
array with weak bonds at the vertical edges as in Fig. (2a) there will be a
linear resistance 
\begin{equation}
R_L(T)\sim 1/L\xi _{\Vert }\sim L^{-1}(T-T_{cI})  \label{eq:3}
\end{equation}
provided there are free boundary conditions at these edges. This can be
realized by uniform current injection. Other methods of current injection
such as injection from superconducting busbars will suppress this effect as
the busbars repel the vortices (charges) from the edges and the linear
resistance will be reduced to $O(L^{-\pi K})$\cite{hn,sim} as in arrays with
strong bonds at the vertical edges.

\begin{figure}[tbp]
\centering\epsfig{file=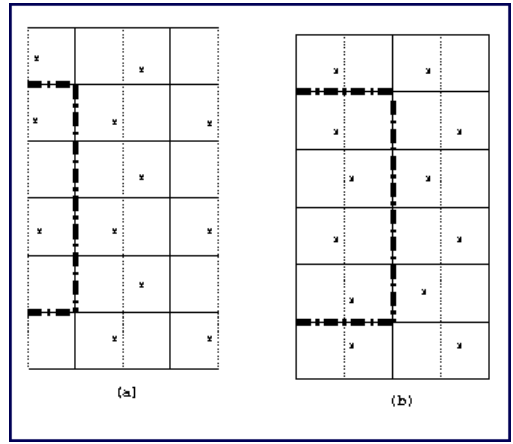,bbllx=0.5cm,bblly=7cm,bburx=17cm,
bbury=22cm,angle=-90,width=8cm}
\caption{(a) Ising domain at array edge with weak bond at edge. (b) Edge
domain for strong bond at edge.}
\label{fig2}
\end{figure}

To check the above qualitative predictions, we have performed simulations on
arrays of both geometries and with both uniform current injection and
injection from superconducting busbars. Also, to avoid the effects of finite
applied current which causes a non linear $IV$ relation in the thermodynamic
limit\cite{ahns,hn}, we have computed the linear resistance from a linear
response expression in terms of equilibrium quantities. Our simulations are
performed using the Langevin dynamical method of Falo {\it et al.}\cite{falo}
. We assume that the junctions on alternate columns have critical currents $
\eta I_0$ and $I_0$ elsewhere and are shunted by equal resistances $R_s$.
Each superconducting grain has a small capacitance $C$ to ground. The
dynamical equations for the phases $\theta _i$ and the voltages $V_i$ of the
grain at site $i$ follow from charge conservation and the Josephson equation

\begin{eqnarray}
d\theta _i/dt &=&2eV_i/\hbar \nonumber \\
CdV_i/dt &=&I_o\sum_{<j>}\eta _{ij}\sin (\theta _j-\theta
_i-A_{ij}) \nonumber \\
&+ & R_s^{-1}\sum_{<j>}(V_j-V_i)  +\sum_{<j>}I_{ij}^{th}
\end{eqnarray}
where $\eta _{ij}=1$ on all horizontal bonds and $\eta _{ij}=\eta ,1$ on
alternating vertical bonds as in Fig .(1) and the sums over $<j>$ are over
the nearest neighbors of site $<i>$. The thermal noise current $I_{ij}^{th}$
on the bond $<ij>$ is Gaussian distributed and obeys the fluctuation
dissipation theorem 
\begin{equation}
<I_{ij}^{th}(t)I_{kl}^{th}(t^{\prime })>=(2T/R_s)(\delta _{ik}\delta
_{jl}-\delta _{il}\delta _{jk})\delta (t-t^{\prime })
\end{equation}
For the columns of grains at either edge of the array labelled by $
i=1,2,\cdots ,L_y$

\begin{eqnarray}
CdV_i^L/dt &=&I+I_0\sum_{<j>}\eta _{ij}\sin (\theta _j-\theta
_i-A_{ij}) \nonumber \\
&+ & R_s^{-1}\sum_{<j>}(V_j-V_i^L) 
 + \sum_{<j>}I_{ij}^{th}  \nonumber \\
CdV_i^R/dt &=&-I+I_0\sum_{<j>}\eta _{ij}\sin (\theta _j-\theta
_i-A_{ij})  \nonumber \\
&+ &R_s^{-1}\sum_{<j>}(V_j-V_i^R) 
+ \sum_{<j>}I_{ij}^{th}
\label{eq:uniform}
\end{eqnarray}
where $I$ is the current per bond injected into each grain on the left
column and extracted from each grain on the right column. For convenience,
we choose system sizes with the number of columns $L_x$ an {\it odd} integer
with free boundary conditions at the left and right hand edges and the
number $L_y$ of horizontal rows an {\it even} integer with periodic boundary
conditions in the vertical direction. We choose to use the Landau gauge
where $A_{ij}^x=0$ on all horizontal bonds, $A_{ij}^y=\pi $ on the odd
numbered vertical columns (first, third, $\cdots $, last) and $A_{ij}^y=0$
on even numbered columns. This is a particularly convenient choice of
lattice size and gauge as it permits an integer number of $2\times 2$ unit
cells and also for a simple formulation when superconducting busbars are
connected to the left and right hand edges of the array by a set of $L_y$
junctions as the phase and voltage on each busbar is then $y$-independent.
The equations governing these phases $\theta _L(t)$, $\theta _R(t)$ and the
voltages $V_L(t)$, $V_R(t)$ are then\cite{sim} 
\begin{eqnarray}
d\theta _{L,R}/dt &=&2eV_{L,R}/\hbar  \nonumber \\
CdV_L/dt &=&I+L_y^{-1}\sum_{i=1}^{i=L_y}[I_0\sin (\theta _i-\theta
_L) \nonumber \\
&+ & R_s^{-1}(V_i-V_L)+I_{iL}^{th}]  \nonumber \\
CdV_R/dt &=&-I+L_y^{-1}\sum_{i=1}^{i=L_y}[I_0\sin (\theta _i-\theta
_R) \nonumber \\
&+ & R_s^{-1}(V_i-V_R)+I_{iR}^{th}]  \label{eq:busbars}
\end{eqnarray}
Here the sums over $i$ are over the $L_y$ sites connected to the busbars.
The mean voltage drop $V$ across the system is given by 
\begin{equation}
V/R_sI_0=(\phi (t_r)-\phi (0))/t_r  \label{eq:Vphase}
\end{equation}
where $\phi (t)=\theta _L(t)-\theta _R(t)$ is the phase difference across
the array at time $t$ and $t_r$ is the run time of a simulation. The
temperature $T$ is measured in units of $\hbar I_0/2e$ and the time $t$ is
in units of $1/\omega _J=(\hbar C/2eI_0)^{1/2}$, the inverse Josephson
plasma frequency. A time step of typically $\Delta t=0.05$ in these units
\cite{sim} was used in the numerical integration. Changing the time step
does not change the results. Most of our simulations have been done on small
systems with $L_x=17$ and $L_y=16$ with periodic boundary conditions in the $
y$ direction. In any event, we found that boundary effects in the transverse 
$y$ direction are not significant to the accuracy of our simulations.

The $IV$ relation is shown in Figs.(3) and (4) for the anisotropy parameter $
\eta =0.5$ at different temperatures $T$ with the current $I$ applied
perpendicular to the weak bonds using the method of uniform injection
described by Eq.(\ref{eq:uniform}) with periodic boundary conditions in the
transverse $y$ direction. For the array of Fig. (2a), the onset of linear
Ohmic behavior $V=R_{L}I$ is consistent with the data for $T\geq 0.2$ which
is close to the estimate of the Ising critical temperature $T_{cI}\approx
0.18$ for this value of $\eta =0.5$. When the geometry is changed so that
the strong bonds are at the edges as in Fig. (2b), Ohmic behavior is
observed only at higher temperatures $T=0.5-0.6$, which is closer to the
estimate of the $XY$ transition at $T_{cXY}\approx 0.5$ as expected. For $
\eta =2$, as expected from the qualitative arguments, the situation is
reversed and the results of the simulations are shown in Fig. (5) for arrays
with the $\eta$ or strong bonds at the edges as in Fig. (2a) and in Fig. (6)
for arrays of with the $\eta$ bonds not at the edges as in Fig. (2b). In the
latter case, the onset of linear dissipation is observed at $T\approx 0.4$,
close to $T_{cI}$ while in the former case at $T\approx 0.6$, closer to the $
XY$ transition. Again this agrees with our qualitative arguments.

\begin{figure}[tbp]
\centering\epsfig{file=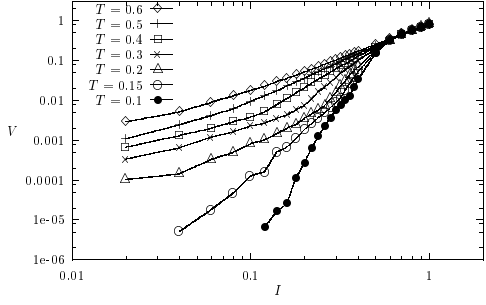,bbllx=3cm,bblly=10cm,bburx=18cm,
bbury=20cm,width=8cm}
\caption{$IV$ characteristics for array with weak edge bonds as in Fig.
(2a). $\eta =0.5, L_{x}=17, L_{y}=16$ and periodic BC in $y$ direction.}
\label{fig3}
\end{figure}

\begin{figure}[tbp]
\centering\epsfig{file=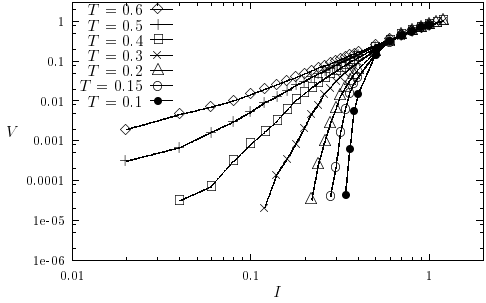,bbllx=3cm,bblly=10cm,bburx=18cm,
bbury=20cm,width=8cm}
\caption{$IV$ characteristics for array with strong edge bonds as in Fig.
(2b). $\eta =0.5$, $L_{x}=17, L_{y}=16$ and periodic BC in $y$ direction.}
\label{fig4}
\end{figure}

\begin{figure}[tbp]
\centering\epsfig{file=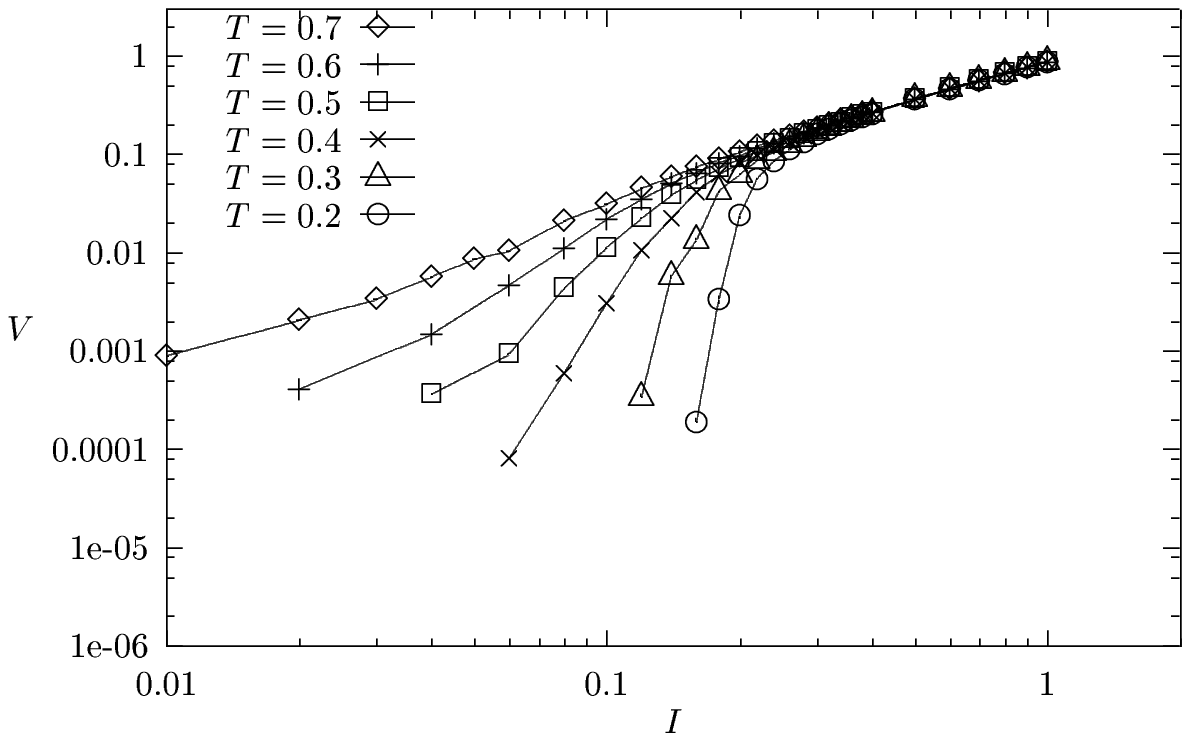,bbllx=3cm,bblly=10cm,bburx=18cm,
bbury=20cm,width=8cm}
\caption{$IV$ characteristics for array with strong edge bonds. $\eta =2,
L_{x}=17, L_{y}=16$ and periodic BC in $y$ direction.}
\label{fig5}
\end{figure}

\begin{figure}[tbp]
\centering\epsfig{file=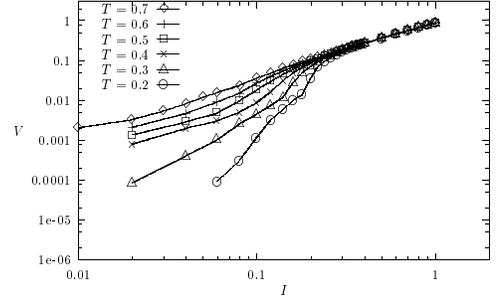,bbllx=3cm,bblly=10cm,bburx=18cm,
bbury=20cm,width=8cm}
\caption{$IV$ characteristics for array with weak edge bonds. $\eta =2,
L_{x}=17, L_{y}=16$ and periodic BC in $y$ direction.}
\label{fig6}
\end{figure}

When the applied current $I$ is finite, the $IV$ relation always has a
non-linear contribution. The resistance, defined by $R=V/I$, is proportional
to $I^{-\pi K_R(T)}$ for $T\leq T_{cI}$ in the thermodynamic limit which may
obscure the small predicted linear resistance of $O(1/L)$. To obtain a more
definitive signature which is free of these non-linear finite current
effects, one can define the linear resistance $R_L$ by a Kubo formula in
terms of an equilibrium voltage-voltage correlation function at $I=0$ 
\begin{equation}
R_L=(1/2T)\int_{-\infty }^{+\infty }dt<V(t)V(0)>  \label{eq:kubo}
\end{equation}
where $V(t)$ is the total size-dependent voltage across the array. The
results of the dynamical simulation using the Kubo expression of
Eq.(\ref{eq:kubo})
at $I=0$ are shown in Fig. (7) together with those from a simulation with $
I/I_0=0.02$, the lowest accessible current, and defining $R_L=V/I$. This is
always larger than the value obtained from Eq.(\ref{eq:kubo}) because of the
non-linear contribution when $I>0$. This can be regarded as confirmation of
our arguments that the onset of linear dissipation at $T_{cI}$ is not an
artifact but a real effect caused by the Ising transition and may be
observable by experiment. Our results for arrays of size $L_x=17,\ L_y=16$
with $\eta =0.5$ and periodic boundary conditions in the $y$ direction are
summarized in Fig. (8) where we show the linear resistance $R_L(T)$ computed
from the Kubo expression of Eq.(\ref{eq:kubo}) for arrays geometries of Fig.
(2a) and Fig. (2b). From this it is clear that there is an onset of linear
resistance around $T_{cI}$ when the weak bonds are at the edges and at a
higher temperature $T\approx T_{cXY}$ when strong bonds are at the edges.
Our results for the phase diagram in the $(T,\eta )$ plane are shown in Fig.
(9) where we plot $T_{cI}(\eta )$ determined by purely equilibrium Monte
Carlo simulations and also by the onset of $R_L$ for arrays of Fig. (2a)
from Eq.(\ref{eq:kubo}). To the accuracy of our simulations, the two methods
agree providing additional evidence to support our qualitative arguments. As
a final check, we performed a series of simulations at finite $I$ on larger
systems up to $L=200$ to check the prediction $R_L\sim 1/L$ of Eq.(\ref{eq:3}
) for $T>T_{cI}$ and in Fig. (10) we show $VL/I$ for several values of $L$
and we see that this is consistent with being an $L$-independent constant
although the errors are fairly large.

\begin{figure}[tbp]
\centering\epsfig{file=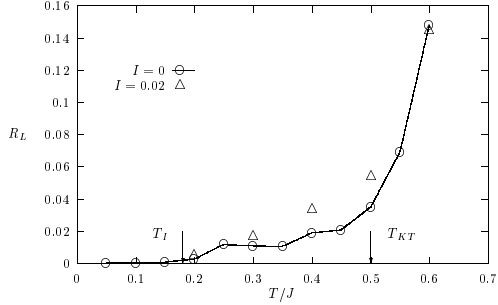,bbllx=3cm,bblly=10cm,bburx=18cm,
bbury=22cm,width=8cm}
\caption{Linear resistance for array with weak edge bonds, $\eta =0.5,
L_{x}=17, L_{y}=16$ and periodic BC in $y$ direction, as estimated from the
Kubo formula of Eq.(\ref{eq:kubo}) at $I=0$ and from $V/I$ at $I=0.02$}
\label{fig7}
\end{figure}

\begin{figure}[tbp]
\centering\epsfig{file=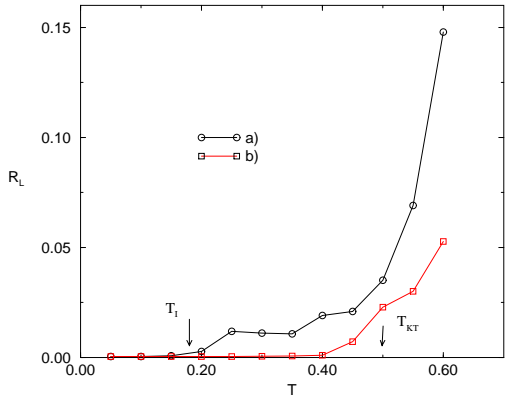,bbllx=1cm,bblly=1cm,bburx=22cm,
bbury=18cm,width=8cm}
\caption{Linear resistance $R_{L}$ as a function of $T$ from Kubo formula of
Eq.(\ref{eq:kubo}) at $I=0$ for array with $L_{x}=17, L_{y}=16, \eta =0.5$
in Fig. 2 with weak edge bonds (a) and with strong edge bonds (b). Note that 
$R_{L}(T)>0$ for $T_{cXY}>T>T_{cI}$ for weak edge bonds and $T>T_{cXY}$ for
strong edge bonds.}
\label{fig8}
\end{figure}

\begin{figure}[tbp]
\centering\epsfig{file=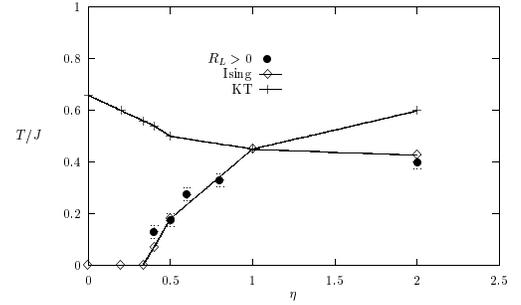,bbllx=3cm,bblly=10cm,bburx=18cm,
bbury=22cm,width=8cm}
\caption{Onset temperature $T$ for $R_{L}(T)>0$ from Kubo formula Eq.(\ref
{eq:kubo}) as function of $\eta$ denoted by solid dots. Ising and $KT$
transition temperatures from equilibrium Monte Carlo simulations are open
diamonds and pluses, respectively.}
\label{fig9}
\end{figure}

The temperatures at which the onset of linear resistance occurs is
consistent with the Ising critical temperature and happens for arrays of
Fig. (2a) when $\eta <1$ and Fig. (2b) when $\eta >1$. This suggests that
the linear resistance observed when $T_{cI}<T<T_{cXY}$ is a result of both
Ising disorder and free boundary conditions at the array edges allowing for
motion of fractional charges induced by the domains, as predicted by
qualitative arguments. The presence of weak bonds at the edges of the arrays
of Fig. (2a) for $\eta <1$ and of Fig. (2b) for $\eta >1$ is not essential
as changing the strength of the edge bonds does not change the scenario in
any qualitative way but only quantitatively. The determining factor is that
the periodicity of the Ising domain walls is two lattice spacings which is a
consequence of the energy of a domain wall on a weak bond being larger than
on a strong bond for $\eta <1$ and the reverse for $\eta >1$. When the edges
correspond to a weak bond, domains at edges have width $n+1/2$ unit cells
which implies that there are free fractional charges at the ends of these
domains in thermal equilibrium and these are free to move under the
influence of an applied current. The strength of the edge bonds does not
affect the argument in any essential way except to alter the magnitude of
the dissipation. We have checked that the dissipation onset at $T_{cI}$ is
in fact an edge effect by performing simulations in a busbar geometry where
the current is injected and extracted by attaching the edges of the array to
superconducting busbars\cite{sim} described by Eq.(\ref{eq:busbars}). In
this geometry, fractional charges (vortices) are repelled from the edges and
domain formation is suppressed, thus minimizing or completely suppressing
edge contributions to the dissipation. The $IV$ relation for such an array
of Fig. (2a) with $\eta =0.5$ is shown in Fig. (11) for the current $I$
normal to the weak bonds and in Fig. (12) for $I$ parallel to the weak
bonds. There is good agreement between the $IV$ relations of Fig. (11) and
Figs.(4,5) in which there are strong bonds at the edges and where we expect
there to be no linear contribution. One also observes that the slopes of the 
$IV$ curves in the busbar geometry are independent of the direction of $I$
as the slopes in Figs.(11) and (12) are the same for the same temperature $T$
. This slope $a(T)=\log V/\log I$ is plotted in Fig. (13) together with that
deduced from the helicity modulus \cite{eikmans} $\gamma =(\gamma _x\gamma
_y)^{1/2}$ by \cite{hn} $a(T)=\pi \gamma /T$. The agreement between the
different methods is by no means perfect but we believe the numerical
support for our qualitative predictions is more than adequate to demonstrate
their validity.

\begin{figure}[tbp]
\centering\epsfig{file=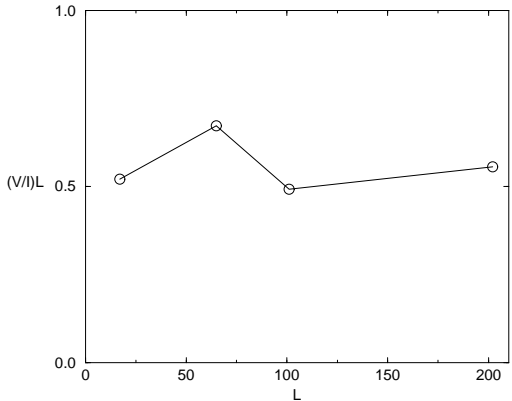,bbllx=1cm,bblly=1cm,bburx=22cm,
bbury=20cm,width=8cm}
\caption{Plot of $(V/I)L$ at $T=0.4, I=0.02$ against array size $L$.}
\label{fig10}
\end{figure}

\begin{figure}[tbp]
\centering\epsfig{file=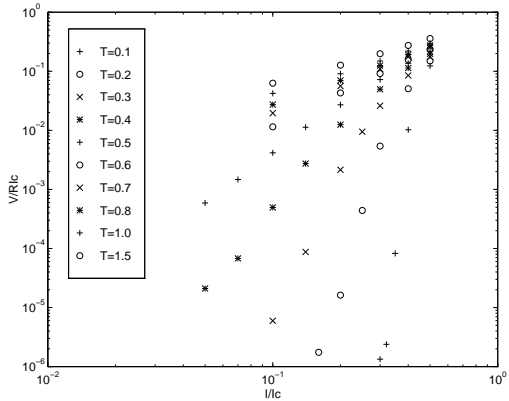,bbllx=1cm,bblly=6cm,bburx=22cm,
bbury=20cm,width=8cm}
\caption{$IV$ characteristics of arrays with weak bonds at edges in busbar
geometry of size $L_{x}=17, L_{y}=16, \eta =0.5$ with current in $x$
direction, normal to weak bonds.}
\label{fig11}
\end{figure}

\begin{figure}[tbp]
\centering\epsfig{file=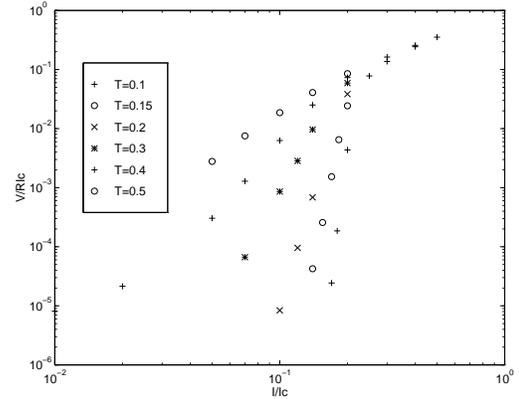,bbllx=1cm,bblly=6cm,bburx=22cm,
bbury=20cm,width=8cm}
\caption{$IV$ characteristics of array with $L_{x}=17, L_{y}=16, \eta =0.5$
and current in $y$ direction along weak bonds.}
\label{fig12}
\end{figure}

\begin{figure}[tbp]
\centering\epsfig{file=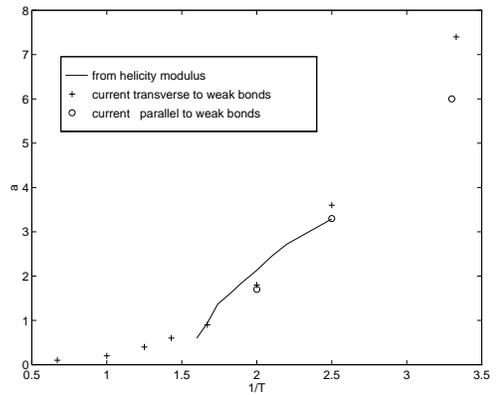,bbllx=1cm,bblly=6cm,bburx=22cm,
bbury=20cm,width=8cm}
\caption{Exponent $a(T)$ from $IV$ curves and helicity modulus.}
\label{fig13}
\end{figure}

The conclusion we reach is that the low $T$ Ising transition in anisotropic
arrays of Josephson junctions does have an effect which may be observable in
zero frequency transport measurements. This effect is the onset of a linear
resistance at $T_{cI}$ in arrays of the appropriate geometry. Although this
is an edge effect with $R_L(T)\sim (T-T_{cI})/L$ it should be large enough
to be measurable in an array of reasonable size. This onset is an
unambiguous signature of the low $T$ Ising transition as, in the
thermodynamic limit $R_L(T)L=A(T)\sim (T-T_{cI})$ and bulk finite size
corrections to this will decay very rapidly with increasing $L$ and should
be negligible compared to the $1/L$ contribution discussed here. These bulk
finite size effects can be estimated by performing the same measurement with
the current parallel to the weak bonds when the $1/L$ contribution to $R_L$
will be absent, which provides a method of assessing the feasibility of the
proposed measurement. This small edge contribution to the linear resistance
is a real effect which is a signal of the bulk Ising transition but, as is
well known, the interpretation of such measurements is very difficult at
temperatures well below the $KT$ temperature $T_{cXY}$ because of screening
effects\cite{martinoli,martinoli1}. We have not attempted to take such
effects into account in this work but there is a possibility that they may
invalidate our conclusions. Clearly this should be investigated as should
various other effects such as disorder in the junction strengths which will
inevitably be present in a real system, vortex
and domain wall pinning etc. It is not very aesthetically satisfying that one
must resort to an edge effect detectable by a voltage measurement at finite
current as there are limits to the sensitivity of this, especially because a
typical experimental array\cite{pm} is much larger than those of this work
so that the dissipation at the edges will be much smaller than in our
simulations. In fact, in the thermodynamic limit $L\rightarrow\infty$ the
linear dissipation in the temperature range $T_{cI}<T<T_{cXY}$ where the
vortex lattice is melted should vanish. It would be much
more satisfying if some equilibrium bulk quantity would give a signal of the
Ising transition but in view of the very weak signal in the helicity modulus
\cite{eikmans,gk1}, this will be a very difficult effect to observe. 
There is the possibility that a
flux noise measurement similar to that of Shaw {\it et al.}\cite{shaw} may
show a detectable signal of the Ising transition. However, naively this
seems not very hopeful as the noise spectrum basically measures the time
dependent correlation function $<N(t)N(0)>$ where $N(t)$ is the total
vorticity (charge) enclosed by the squid detector. Since the Ising
transition is signalled by the proliferation of domains with fractional
corner charges but with zero net charge, as in Fig. (1b), it is difficult to
see how these can contribute significantly to the correlation function
determining the noise spectrum. However, this is still worth investigating
as it is one of the few remaining possibilities to detect this elusive
transition.

\acknowledgements
This work was supported by the IAE agency/ICTP (E.G.) and by a joint
NSF-CNPq grant (E.G.and J.M.K.). The authors thank H. Pastoriza and P.
Martinoli for correspondence about their experimental results prior to
publication. Some of the computations were
performed at the Theoretical Physics Computing Facility at Brown University.

\end{document}